\documentclass[aps,showpacs,prd,twocolumn]{revtex4}
\usepackage{graphicx}
\usepackage{amssymb}
\usepackage{amsmath}
\usepackage{color}
\usepackage{float}
\usepackage{accents}
\usepackage{graphicx}
\usepackage{graphicx}

\begin{document}

\title{Generalized BPS magnetic monopoles}
\author{R. Casana$^{1}$, M. M. Ferreira Jr.$^{1}$ and E. da Hora$^{1,2}$.}
\affiliation{$^{1}${Departamento de F\'{\i}sica, Universidade Federal do Maranh\~{a}o,
65085-580, S\~{a}o Lu\'{\i}s, Maranh\~{a}o, Brazil.}\\
$^{2}${Grupo de F\'{\i}sica Te\'{o}rica Jayme Tiomno, S\~{a}o Lu\'{\i}s,
Maranh\~{a}o, Brazil.}}

\begin{abstract}
We show the existence of Bogomol'nyi-Prasad-Sommerfield (BPS) magnetic
monopoles in a generalized Yang-Mills-Higgs model which is controlled by two
positive functions, $g\left( \phi ^{a}\phi ^{a}\right) $ and $\ f\left( \phi
^{a}\phi ^{a}\right) $. This effective model, in principle, would describe
the dynamics of the nonabelian fields in a chromoelectric media. We check
the consistency of our\textbf{\ }generalized construction by analyzing an
explicit case ruled by a parameter $\beta $. We also use the well-known
spherically symmetric \textit{Ansatz} to attain the corresponding self-dual
equations describing the topological solutions. The overall conclusion is
that the new solutions behave around the canonical one, with smaller or
greater characteristic length depending on the values of $\beta $.
\end{abstract}

\maketitle

\section{Introduction}

\label{Intro}

In the context of classical field theories, configurations with nontrivial
topology, generically named \textit{topological defects}, arise\ as finite
energy solutions to some nonlinear models. In the standard approach, such
models are usually endowed by a symmetry breaking potential for the matter
self-interaction, since topological defects are known to be formed during
symmetry breaking phase transitions \cite{n5}.

The simplest topological defect is the static kink, which arises in a
(1+1)-dimensional theory containing a single real scalar field \cite{n0}.
Other well-known examples of topological structures are the vortex
configurations, which appear in effective planar models containing a complex
scalar field coupled to an abelian gauge field \cite{n1}, and the magnetic
monopole, which stand for the static solution arising from a
(1+3)-dimensional theory describing the interaction between a real scalar
triplet and non-abelian gauge fields \cite{n3}.

In particular, magnetic monopoles are spherically symmetric configurations
coming from a static Yang-Mills theory endowed with a fourth-order Higgs
potential \cite{Shnir}. In this case, these solutions exhibit no divergences
and possess finite total energy, since they are constrained by a set of
suitable boundary conditions. On the other hand, in the absence of a Higgs
potential, magnetic monopoles arise as the minimal energy configurations of
the corresponding Yang-Mills model \cite{n4}. In this context, such
solutions come from a set of first-order differential equations, and
minimize the total energy of the overall theory.

Furthermore, during the last years, some generalized or effective field
theory models, generically named k-field theories, have been intensively
investigated. The main difference between them and their canonical
counterparts is the presence of nonstandard kinetic terms, which change the
dynamics of the overall system in an exotic way. In fact, such theories have
been used as effective models mainly in Cosmology, with the so-called
k-essence models \cite{n7h} suggesting new insights about{\ }the accelerated
inflationary phase of the universe \cite{n8}. The interesting point is that
the introduction of generalized kinetic terms has important consequences on
the formation of topological structures. Despite the possibility of
achieving nontrivial topologically configurations even in the absence of the
spontaneous symmetry breaking phenomenon \cite{n14a}, some k-field models
endowed with spontaneous symmetry breaking also support topological defects 
\cite{n14b}. In the last case, the resulting solutions can be studied via
the comparison between them and their canonical counterparts, with the
observation that they usually present slight variations on some of their
main features \cite{n14c}.

On the other hand, in the presence of a generalized dynamics, the resulting
model is highly nonlinear, its solutions being quite hard to find, even in
the presence of suitable boundary conditions. In this context, the
development of a consistent first-order framework is quite useful, since new
topological configurations can be found by solving a set of generalized BPS
equations. As it happens in the usual case, these new configurations
minimize the energy of the overall system by saturating its lower bound \cite%
{Bogo}.

Recently, some of us have performed the development of generalized
first-order frameworks regarding several effective field theories \cite{n14d}%
. However, for simplicity, these models were constructed containing only
abelian fields. In the present work, we go further by presenting a
first-order theoretical framework consistent with a \textbf{\ }generalized
Yang-Mills-Higgs model. This model is controlled by two dimensionless
positive functions, $g\left( \phi ^{a}\phi ^{a}\right) $ and $f\left( \phi
^{a}\phi ^{a}\right) $, which change the dynamics of the non-abelian fields
in a nonusual way. Nevertheless, in order to guarantee the self-duality of
the model, $g$ and $f$ \textbf{\ }are related by means of a simple
constraint. As in the usual case, the total energy of the generalized model
is also bounded from below, and the spherically symmetric self-dual
solutions describe generalized BPS magnetic monopoles. Moreover, in the
appropriate limit, our framework leads to the usual one, as expected.

In order to present our results, this work is outlined as follows. In Sec. %
\ref{general}, we introduce the nonstandard Yang-Mills-Higgs model and the
corresponding first-order theoretical framework, including its generalized
BPS equations. Then, instead of recovering the usual case, we introduce an
explicit generalized model, which is controlled by a single real parameter, $%
\beta $. In Sec.\textbf{\ }\ref{numerical2}, we present the numerical
solution of the generalized BPS equations obtained by means of the
relaxation technique. We depict the profiles of the Higgs and non-abelian
gauge fields describing the generalized BPS magnetic monopoles. We also
comment on the main features of these solutions. In Sec. \ref{end}, we
finalize by summarizing our results and stating our perspectives.

%%%%%%%%%%%%%%%%%%%%%%%%%%%

\section{The theoretical framework}

\label{general}

We begin introducing a generalized version of the Yang-Mills-Higgs model
which is defined by the following Lagrangian density\textbf{:} 
\begin{align}
\mathcal{L}& =-\frac{g\left( \phi ^{a}\phi ^{a}\right) }{4}F_{\mu \nu
}^{b}F^{\mu \nu ,b}+\frac{f\left( \phi ^{a}\phi ^{a}\right) }{2}\mathcal{D}%
_{\mu }\phi ^{b}\mathcal{D}^{\mu }\phi ^{b}  \notag \\
& -V\left( \phi ^{a}\phi ^{a}\right) \text{,}  \label{1}
\end{align}%
where all fields, coordinates and parameters are supposed to be
dimensionless (this can be achieved by using the appropriate mass rescaling
transformations). We use standard conventions, including the natural units
system. Here,%
\begin{equation}
F_{\mu \nu }^{a}=\partial _{\mu }A_{\nu }^{a}-\partial _{\nu }A_{\mu
}^{a}+e\epsilon ^{abc}A_{\mu }^{b}A_{\nu }^{c}\text{,}
\end{equation}%
is the usual non-abelian field strength, and%
\begin{equation}
\mathcal{D}_{\mu }\phi ^{a}=\partial _{\mu }\phi ^{a}+e\epsilon ^{abc}A_{\mu
}^{b}\phi ^{c}\text{,}
\end{equation}%
stands for the non-abelian covariant derivative, $\eta ^{\mu \nu }=\left(
+---\right) $ is the metric for the (1+3)-dimensional spacetime, and $%
\epsilon ^{abc}$ is the completely antisymmetric tensor (with $\epsilon
^{123}=+1$). We point out that $g\left( \phi ^{a}\phi ^{a}\right) $ and $%
f\left( \phi ^{a}\phi ^{a}\right) $ are arbitrary functions which change the
dynamics of the overall model in a nonusual way.

It is well-known that, in the usual case (i.e., given $g=f=1$), self-dual
configurations only exist in the absence of the Higgs potential. So, for
simplicity, we\ also assume that their generalized counterparts only exist
for $V\left( \phi^{a}\phi^{a}\right) =0$. Furthermore, we suppose that such
generalized configurations are described by the standard spherically
symmetric \textit{Ansatz}%
\begin{equation}
\phi^{a}=\frac{x^{a}H\left( r\right) }{r}\text{\ \ \ and \ \ }%
A_{i}^{a}=\epsilon_{iak}\frac{x_{k}}{er^{2}}\left( W\left( r\right)
-1\right) \text{,}  \label{a}
\end{equation}
where $r^{2}=x^{a}x_{a}$. Since we are\ searching for static configurations,
we choose $A_{0}^{a}=0$. The profile functions $H\left( r\right) $ and $%
W\left( r\right) $ are supposed to obey the usual finite energy boundary
conditions%
\begin{equation}
H\left( 0\right) =0\text{ \ \ and \ \ }W\left( 0\right) =1\text{,}
\label{cc1}
\end{equation}%
\begin{equation}
H\left( \infty\right) =\mp1\text{ \ \ and \ \ }W\left( \infty\right) =0\text{%
.}  \label{cc2}
\end{equation}
These conditions guarantee not only the existence of finite energy
configurations, but also the breaking of the SO(3) symmetry inherent to the
model (\ref{1}).

In general, the weight functions $g\left( \phi ^{a}\phi ^{a}\right) $ and $%
f\left( \phi ^{a}\phi ^{a}\right) $ can be arbitrarily chosen, so that the
Euler-Lagrange equations for $H\left( r\right) $ and $W\left( r\right) $ are
quite hard to solve, even in the presence of the suitable boundary
conditions (\ref{cc1}) and (\ref{cc2}). In this context, the development of
a consistent self-dual theoretical framework is more than desirable, since
it allows to find a set of first-order differential equations that leads to
finite energy configurations attained by numerical procedures. These
configurations are genuine solutions of the overall model,\ once they
automatically fulfill the nonstandard Euler-Lagrange equations coming from
Lagrangian (\ref{1}).

From now on we focus our attention on the development of a BPS theoretical
framework consistent with the nonstandard model (\ref{1}). In order to
perform it, we follow the usual approach, observing\textbf{\ }that the
first-order equations come from the minimization of the total energy of the
system, given by the energy-momentum tensor zero-zero component. In the
present case, such tensor reads%
\begin{equation}
T_{\lambda\rho}=f\mathcal{D}_{\lambda}\phi^{a}\mathcal{D}_{\rho}\phi
^{a}-gF_{\mu\lambda}^{a}F^{a,\mu}{}_{\rho}-\eta_{\lambda\rho}\mathcal{L}%
\text{,}
\end{equation}
from which one gets the nonstandard energy density (already written in terms
of $H\left( r\right) $ and $W\left( r\right) $)%
\begin{align}
\varepsilon & =\frac{g}{e^{2}r^{2}}\left( \left( \frac{dW}{dr}\right) ^{2}+%
\frac{\left( 1-W^{2}\right) ^{2}}{2r^{2}}\right)  \notag \\
& +f\left( \frac{1}{2}\left( \frac{dH}{dr}\right) ^{2}+\left( \frac {HW}{r}%
\right) ^{2}\right) \text{,}  \label{de}
\end{align}
where $g=g\left( H\right) $ and $f=f\left( H\right) $. It is clear that such
functions must be positive in order to avoid problems with the energy of the
model.

Given the \textit{Ansatz} (\ref{a}), we point out that this model only
engenders self-dual solutions when $g$ and $f$ are related to each other\ as%
\begin{equation}
g\left( H\right) =\frac{1}{f\left( H\right) }\text{.}  \label{v}
\end{equation}
In this case, the energy density (\ref{de}) can be rewritten in the form%
\begin{align}
\varepsilon & =\frac{f}{2}\left( \frac{dH}{dr}\pm\frac{1-W^{2}}{er^{2}f}%
\right) ^{2}+\frac{1}{e^{2}r^{2}f}\left( \frac{dW}{dr}\mp efHW\right) ^{2} 
\notag \\
& \mp\frac{1}{er^{2}}\frac{d}{dr}\left( H\left( 1-W^{2}\right) \right) \text{%
,}
\end{align}
whose minimization leads to the following first-order equations:%
\begin{equation}
\frac{dH}{dr}=\mp\frac{1-W^{2}}{er^{2}f}\text{,}  \label{bps1}
\end{equation}%
\begin{equation}
\frac{dW}{dr}=\pm efHW\text{.}  \label{bps2}
\end{equation}
Relations (\ref{bps1}) and (\ref{bps2}) are the self-dual (BPS)\ equations
of the model (\ref{1}).\ After implementing these equations, the BPS\ energy
density becomes%
\begin{equation}
\varepsilon_{bps}=\mp\frac{1}{er^{2}}\frac{d}{dr}\left( H\left(
1-W^{2}\right) \right) ,  \label{buc}
\end{equation}
and the total energy of the solutions is given by%
\begin{equation}
E_{bps}=4\pi\int r^{2}\varepsilon_{bps}dr=\frac{4\pi}{e}\text{,}  \label{te}
\end{equation}
whenever the boundary conditions (\ref{cc1}) and (\ref{cc2}) are fulfilled.

In summary, for a given positive function $f\left( H\right) $, the
first-order equations{\ (\ref{bps1}) and (\ref{bps2}) }must be numerically
solved in accordance with the finite energy boundary conditions {(\ref{cc1})
and (\ref{cc2})}. The self-dual solutions achieved in this way describe the
generalized BPS magnetic monopoles arising from this nonstandard
Yang-Mills-Higgs model {(\ref{1}), }with total energy given by (\ref{te})
and energy density by (\ref{buc}). However, it is worthwhile to point out
that the generalized first-order framework presented in this letter only
holds in the absence of the potential $V\left( \phi^{a}\phi^{a}\right) $.
Furthermore, $g$ and $f$ \ have to obey the constraint {(\ref{v})}. On the
other hand, for a nonvanishing potential, or in the absence\textbf{\ }of the
relation {(\ref{v})}, such framework does not hold anymore, and the
corresponding BPS monopoles can not be achieved.

The usual results are trivially recovered by setting $f=1$. In order to show
how this generalized framework works, we adopt the following generalization
function: 
\begin{equation}
f\left( H\right) =\left( H^{2}+1\right) ^{\beta}\text{,}  \label{nw}
\end{equation}
where $\beta$ is some real number; here, $\beta=0$ leads us back to the
canonical model. Now, given {(\ref{nw})}, the self-dual expressions {(\ref%
{bps1}) and (\ref{bps2}) }become%
\begin{equation}
\frac{dH}{dr}=\mp\frac{1-W^{2}}{er^{2}\left( H^{2}+1\right) ^{\beta}}\text{,}
\label{nbps1}
\end{equation}%
\begin{equation}
\frac{dW}{dr}=\pm eHW\left( H^{2}+1\right) ^{\beta}\text{,}  \label{nbps2}
\end{equation}
which must be solved respecting the conditions {(\ref{cc1}) and (\ref{cc2})}.

In the next Section, we solve the first order equations {(\ref{nbps1}) and (%
\ref{nbps2}) }by means of the relaxation technique for different values of $%
\beta $. Then, we plot not only the numerical results for $H\left( r\right) $
and $W\left( r\right) $, but also those for the BPS energy density {(\ref%
{buc})}, and for $r^{2}\varepsilon _{bps}$ (the integrand of {(\ref{te})}).
We also comment on the main features of the new solutions.

%%%%%%%%%%%%%%%%%%%%%%%%

\section{Numerical results}

\label{numerical2}

Now, we focus our attention on the examination of the profiles of the
generalized BPS\ solutions. Thus, we numerically solve the first-order
equations (\ref{nbps1}) and (\ref{nbps2}) obeying the finite energy boundary
conditions (\ref{cc1}) and (\ref{cc2}). Here, for simplicity, we choose $e=1$%
, and consider only the lower sign in\ {(\ref{cc2}), (\ref{buc}), (\ref%
{nbps1}) and (\ref{nbps2})}. Then, we numerically solve the self-dual system
by means of the relaxation technique, for different values of the real
parameter $\beta $.

Note that $\beta =0$ leads us back to the usual model, whose self-dual
solutions (already written according our conventions) can be attained
analytically as%
\begin{equation}
H\left( r\right) =\frac{1}{\tanh r}-\frac{1}{r}\text{,}
\end{equation}%
\begin{equation}
W\left( r\right) =\frac{r}{\sinh r}\text{.}
\end{equation}%
The numerical solutions we found for $H\left( {r}\right) $\ and $W\left( {r}%
\right) $\ are depicted\ in Figs. 1 and 2, for $\beta =-2$ (dot-dashed green
line), $\beta =-1$ (dashed blue line), $\beta =1$ (long-dashed orange line)
and $\beta =2$ (dotted red line). The usual (analytical) profiles are also
shown (solid black line), for comparison. Moreover, we also plot the
corresponding solutions for the BPS energy density (\ref{buc}) and for $%
r^{2}\varepsilon _{bps}$ (the integrand of (\ref{te}))\ in Figs. 3 and 4. 
\begin{figure}[tbp]
\centering\includegraphics[width=8.5cm]{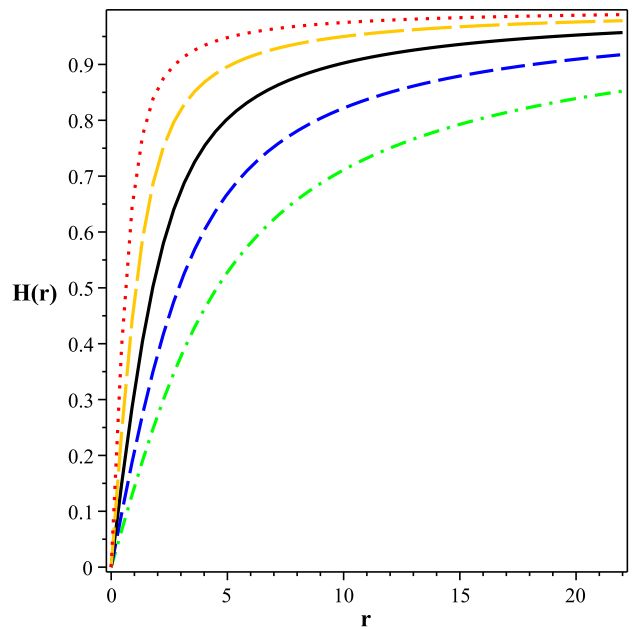}
\par
\vspace{-0.3cm}
\caption{Solutions to $H\left( r\right) $ for $\protect\beta =-2$
(dot-dashed green line), $\protect\beta =-1$ (dashed blue line), $\protect%
\beta =0$ (usual case, solid black line), $\protect\beta =1$ (long-dashed
orange line) and $\protect\beta =2$ (dotted red line).}
\end{figure}

In Figure 1, we present the numerical results regarding the profile of the
function $H\left( r\right) $. In this case, we clearly see that the
nonstandard solutions turn out depicted around the usual counterpart,
behaving in the same general way, but changing the width of the defect.

We point out that different solutions engender different characteristic
lengths. In particular, the solutions arising for $\beta <0$ reach the
asymptotic condition $H\left( \infty \right) =1$ slower than the standard
profile. In this sense, these solutions possess a greater characteristic
length, so that the corresponding bosons mediate long-ranged interactions.
On the other hand, the solutions related to $\beta >0$ reach the saturation
region faster, exhibiting smaller characteristic lengths, which is
associated with small-ranged interactions. 
\begin{figure}[tbp]
\centering\includegraphics[width=8.5cm]{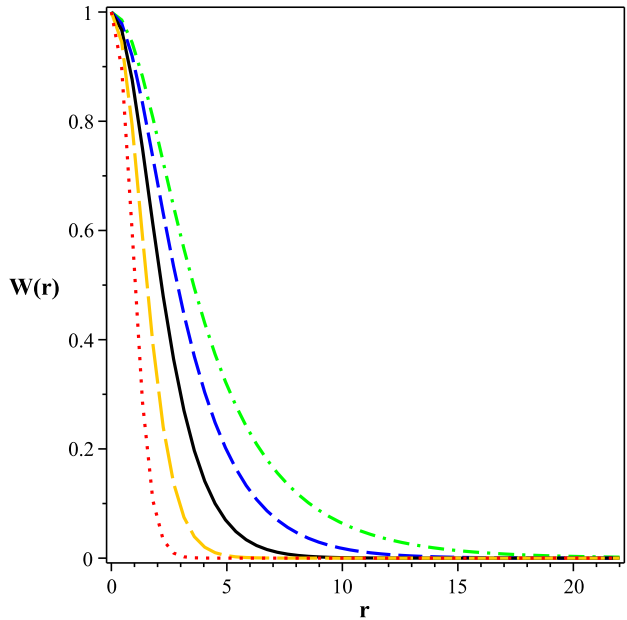}
\par
\vspace{-0.3cm}
\caption{Solutions to $W\left( r\right) $. Convention as in FIG. 1.}
\end{figure}
\begin{figure}[tbp]
\centering\includegraphics[width=8.5cm]{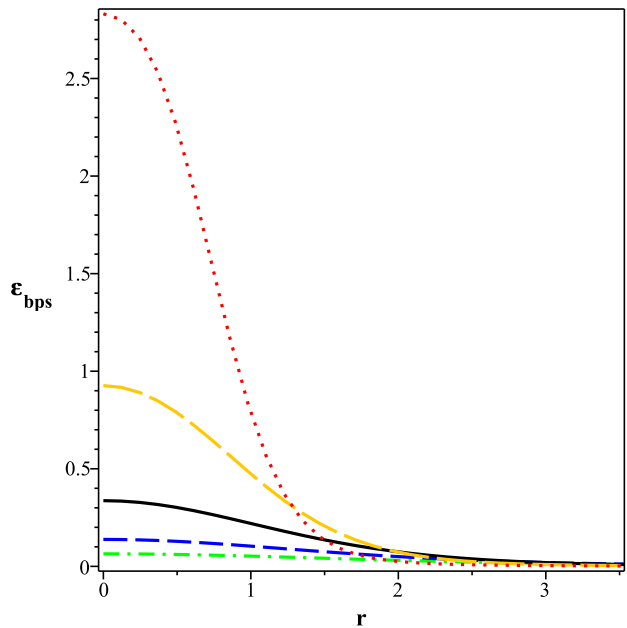}
\par
\vspace{-0.3cm}
\caption{Solutions to $\protect\varepsilon _{bps}$. Convention as in FIG. 1. 
}
\end{figure}

In Fig. 2, we plot the solutions for $W\left( r\right) $. Again, the
nonstandard solutions behave as the canonical one,\ exhibiting variations
concerned with the defect width and the characteristic length. Such
variations are controlled by the real parameter $\beta $ in the same way as
before: for $\beta <0$, the nonusual solutions engender the greater
characteristic lengths, since they reach the asymptotic condition $W\left(
\infty \right) =0$ more slowly, being associated with long-ranged mediating
bosons. Furthermore, for and increasing positive $\beta $\textbf{, }the
solutions reach the saturation region faster, exhibiting smaller and smaller
characteristic lengths, with the mediating bosons yielding small-ranged
interactions.

The numerical profiles found for the BPS energy density $\varepsilon _{bps}$
(\ref{buc}) are depicted in\ Fig. 3,\ revealing that such solutions are
lumps centered at $r=0$. Also, they vanish monotonically as $r$ goes to $%
\infty $ (in fact, $\varepsilon _{bps}\left( r\rightarrow \infty \right)
\rightarrow 0$ arises in a rather natural way from the asymptotic boundary
conditions (\ref{cc2})). Here, the parameter $\beta $ plays a different
role, since it changes not only the characteristic lengths of the
corresponding solutions, but also their amplitudes. In this sense, the
solutions corresponding to $\beta <0$\ $\left( \beta >0\right) $\ achieve
the smaller (larger) amplitudes. 
\begin{figure}[tbp]
\centering\includegraphics[width=8.5cm]{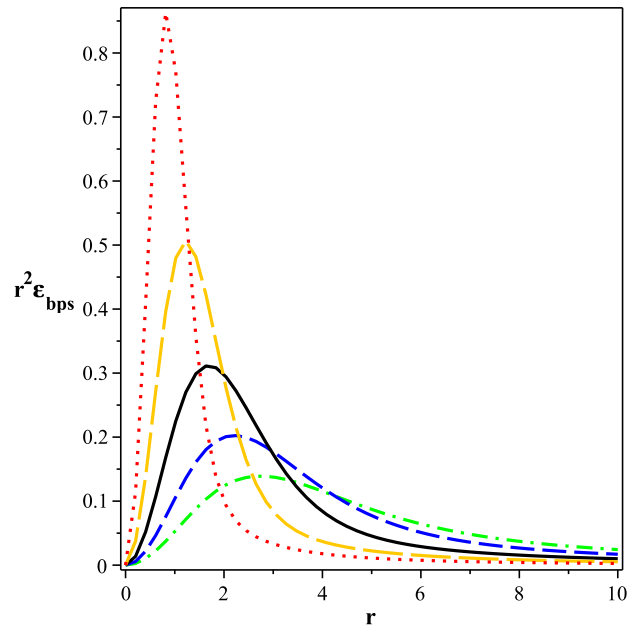}
\par
\vspace{-0.3cm}
\caption{Solutions to $r^{2}\protect\varepsilon _{bps}$. Convention as in
FIG. 1.}
\end{figure}

Finally, we depict the solutions for $r^{2}\varepsilon_{bps}$, i.e., the
integrand of (\ref{te}); see Fig. 4. In this case, the solutions are rings
centered at $r=0$. Moreover, the points of larger amplitudes are located at
some finite distance $R$ from the origin (in this sense, $R$ stands for the
"radius" of the ring), such amplitudes being controlled by $\beta$ in the
same way as before. Here, we point out the existence of an interesting
compensatory effect: the solutions reaching the greater amplitudes spread
over smaller distances, and vice-versa. As a consequence, different
solutions enclose the same area (equal to the unity, according our
conventions), and the corresponding configurations achieve the same total
energy; see (\ref{te}).

%%%%%%%%%%%%%%%%

\section{Ending comments}

\label{end}

In this work, we have presented a nonstandard first-order framework
consistent with a generalized Yang-Mills-Higgs model (\ref{1}), such model
being controlled by two dimensionless functions, $g\left( \phi ^{a}\phi
^{a}\right) $ and $f\left( \phi ^{a}\phi ^{a}\right) $, which change the
dynamics of the non-abelian fields in a non-usual way. In order to avoid
problems with the energy of the system, these functions are supposed to be
positive. We point out that the self-duality of the overall model only holds
when $g$ and $f$ are related to each other by a simple constraint; see (\ref%
{v}). Also, it is worthwhile to note that there is no additional constraint
to be imposed on $f$.

The generalized first-order framework was developed in a general way. So, in
order to verify the consistency of our construction, {we have introduced an}
explicit example {controlled by a single real parameter }${\beta }${; see (%
\ref{nw}). }Then, we have considered spherically symmetric self-dual
configurations, the non-abelian fields being described by the standard
static \textit{Ansatz} (\ref{a}). Moreover, the profile functions ${H}\left( 
{r}\right) ${\ and } ${W}\left( {r}\right) $ were supposed to obey the usual
finite energy boundary conditions given by (\ref{cc1}) and (\ref{cc2}).

The resulting first-order equations were solved numerically by means of the
relaxation technique, and the self-dual solutions we found were plotted in
Figs. 1, 2, 3 and 4, for different values of $\beta $. The standard
analytical solutions (achieved for $\beta =0$) were also plotted, for
comparison. The overall conclusion is that the nonstandard solutions behave
in the same general way the usual one does, the main difference being slight
variations on the amplitudes and on the characteristic lengths of the new
solutions. In addition, {we have identified the way such variations are
controlled by the real parameter }$\beta ${.}

Recently, some of us have performed detailed investigations addressing
generalized self-dual frameworks for abelian models, attaining their
respective first-order solutions \cite{n14d}. This way, the present letter
is an extension of those works to the non-abelian context. As for future
investigations, interesting issues including the supersymmetric extension of
the non-abelian model (\ref{1}), and the search for its topological
structures, are now under consideration.

The authors thank Caroline dos Santos for useful comments about this work.
We are also grateful to CAPES, CNPq and FAPEMA (Brazil) for partial
financial support.

\end{document}